\documentclass[conference]{IEEEtran}
\IEEEoverridecommandlockouts
\usepackage{float}
\usepackage{cite}
\usepackage{hyperref}
\hypersetup{
    colorlinks=true,
    linkcolor=blue,
    filecolor=magenta,      
    urlcolor=cyan,
    pdftitle={Overleaf Example},
    pdfpagemode=FullScreen,
    }

\usepackage{amsmath,amssymb,amsfonts}
\usepackage{algorithmic}
\usepackage{graphicx}
\usepackage{textcomp}
\usepackage{xcolor}
\def\BibTeX{{\rm B\kern-.05em{\sc i\kern-.025em b}\kern-.08em
    T\kern-.1667em\lower.7ex\hbox{E}\kern-.125emX}}

\begin{document}

\title{FLAG: \textbf{F}inancial \textbf{L}ong Document Classification via \textbf{A}MR-based \textbf{G}NN}

\author{\IEEEauthorblockN{Bolun (Namir) Xia, Aparna Gupta, Mohammed J. Zaki}
\IEEEauthorblockA{
\textit{Rensselaer Polytechnic Institute, Troy, NY, USA} \\
xiabolun@gmail.com, guptaa@rpi.edu, zaki@cs.rpi.edu
}
}
\maketitle

\begin{abstract}
The advent of large language models (LLMs) has initiated much research into their various financial applications. However, in applying LLMs on long documents, semantic relations are not explicitly incorporated, and a full or arbitrarily sparse attention operation is employed. In recent years, progress has been made in Abstract Meaning Representation (AMR), which is a graph-based representation of text to preserve its semantic relations. Since AMR can represent semantic relationships at a deeper level, it can be beneficially utilized by graph neural networks (GNNs) for constructing effective document-level graph representations built upon LLM embeddings to predict target metrics in the financial domain. We propose FLAG: {\bf F}inancial {\bf L}ong document classification via {\bf A}MR-based {\bf G}NN, an AMR graph based framework to generate document-level embeddings for long financial document classification. We construct document-level graphs from sentence-level AMR graphs, endow them with specialized LLM word embeddings in the financial domain, apply a deep learning mechanism that utilizes a GNN, and examine the efficacy of our AMR-based approach in predicting labeled target data from long financial documents. Extensive experiments are conducted on a dataset of quarterly earnings calls transcripts of companies in various sectors of the economy, as well as on a corpus of more recent earnings calls of companies in the S\&P 1500 Composite Index. We find that our AMR-based approach outperforms fine-tuning LLMs directly on text in predicting stock price movement trends at different time horizons in both datasets. Our work also outperforms previous work utilizing document graphs and GNNs for text classification.
\end{abstract}

\section{Introduction}
Textual data is an important qualitative source of information in the financial domain. Financial reports can provide valuable signals for a firm's future performance, since these reports usually contain forward-looking plans and strategies, which may not be fully captured in their financial statements. Since textual data provides greater insights into firm performance, various methods have been utilized for transforming these textual reports into numerical representations, in order to define effective features for predicting target variables such as temporal price trends, that are of value to investors.

Despite the progress that has been made in recent years, especially in the sphere of language models (LMs), there still remains the challenge of long documents whose lengths usually exceed the maximum context length of LMs. Even with longer context LLM, learning good representations of documents is still quite difficult: a recent benchmark work on Q\&A tasks in the financial domain demonstrated that even big LLMs such as \cite{openai2024gpt4} have difficulties in answering questions correctly based on specific corpora of financial documents~\cite{islam2023financebench}. In addition, with transformer-based methods, semantic relations between word entities are usually constructed arbitrarily, either with full attention where each word attends to every other word, or sparse attention, where attentions between words are set up arbitrarily, such as sliding window attention or randomized attention.

We propose {\bf F}inancial {\bf L}ong document classfication via {\bf A}MR-based {\bf G}NNs (FLAG), that learns effective document-level embeddings based on specialized LM word embeddings in the finance domain through AMR \cite{banarescu2013abstract}, which is a graph representation of text that preserves semantic relations. The unique feature of AMR graphs is that they are abstracted representations of text capable of capturing the semantic meaning of sentences, rather than just verbatim word sequences. Hence, words, phrases and sentences that have the same meaning, but differ in wording or spelling, usually result in the same AMR representation. As such, AMR is more semantically detailed and represents deeper meaningful relations between semantic concepts. 

\begin{figure}[!ht]
\centering
  \includegraphics[width=3in]{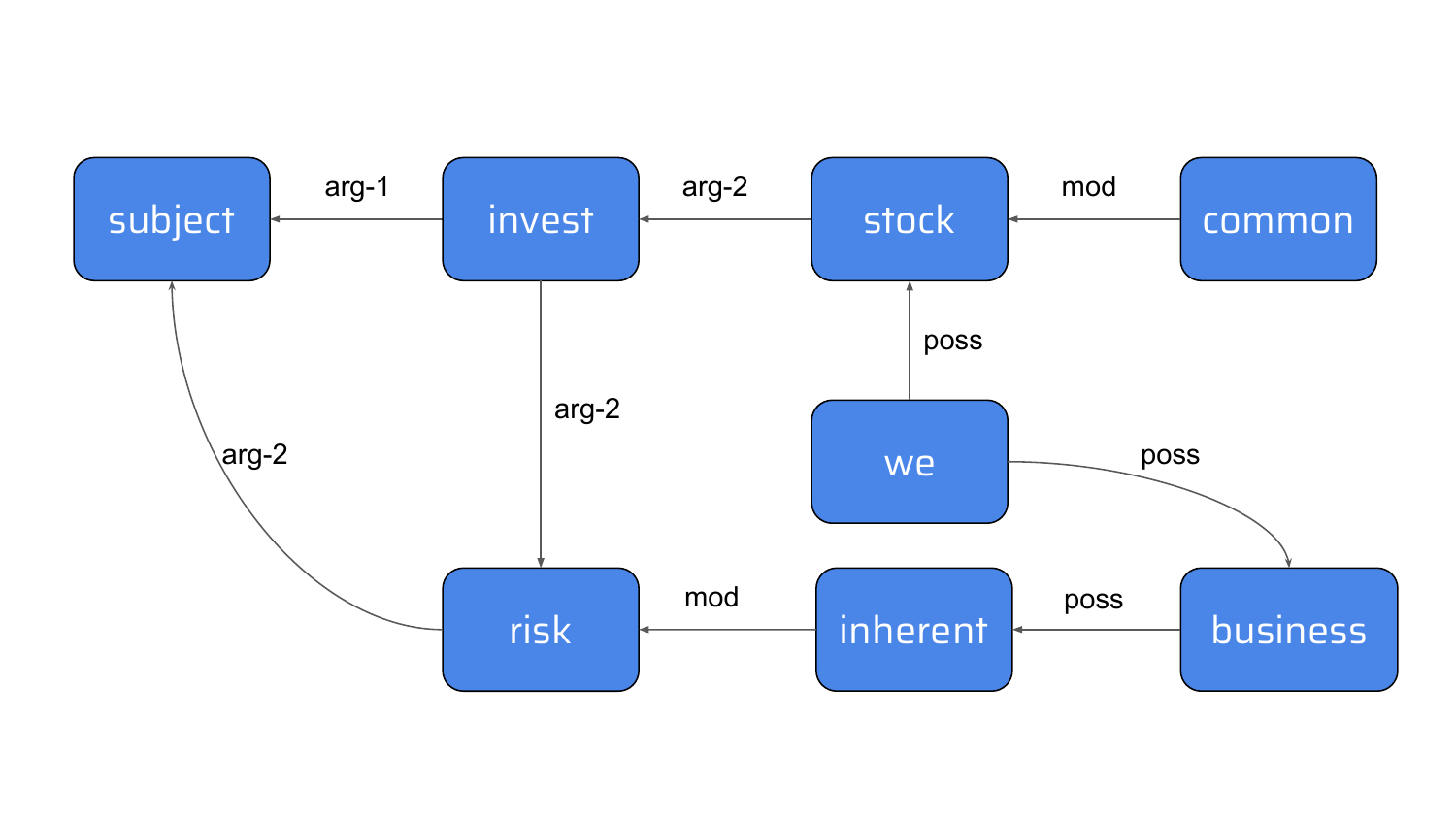}
  \caption{An example of the AMR graph for the sentence: {\em an investment in our common stock is subject to risks inherent to our business}.}
  \label{fig:example-AMR}
\end{figure}

In order to demonstrate the abstracting nature of AMR graphs, in Figure \ref{fig:example-AMR}, we show the AMR graph of a sample sentence: {\em an investment in our common stock is subject to risks inherent to our business}. As we can see, the AMR  graph is an abstracted representation of semantics. It can identify meaningful concepts within a sentence, such as ``invest'' and ``subject'', and it extracts the semantic relations between them. Moreover, we see that the two instances of ``our'' in the original sentence have both been abstracted as the concept of ``we'', and the node ``we'' has possessive relations with ``stock'' and ``business.''

Our approach utilizes the transition AMR parser \cite{naseem-etal-2019-rewarding} to transform each sentence of a document into an AMR graph with alignment of each node with its corresponding word in the sentence. The sentence-level graphs are then aggregated using a hierarchical approach that utilizes both document-level and sentence-level virtual nodes. We initialize each node (or word) with its contextual embedding using FinBERT \cite{yang2020finbert}, a specially trained LM in the finance domain. On the document-graph thus constructed, we apply GATv2 \cite{GATv2}, a recent GNN that employs dynamic attention mechanism. Finally, we take the embedding of the document virtual node as the final representation for the document, and use it for the downstream classification task. While we choose FinBERT as the base LM for FLAG, other models such as BloombergGPT \cite{wu2023bloomberggpt}, or open-source models such as FinGPT \cite{yang2023fingpt} can also be used.

In summary, our contributions are:
\begin{itemize}
    \item We propose and implement an AMR-based deep learning framework for classification tasks geared towards long financial documents. Our FLAG approach constructs novel document-level AMR graphs from sentence-level AMR graphs and uses a GNN to learn effective document-level representations.
    \item We perform an extensive set of experiments on two collections of earnings calls for companies from different sectors of the economy and from the S\&P 1500 Composite Index to show that FLAG outperforms previous methods in predicting stock price movement trends for different time horizons, thereby achieving state-of-the-art (SOTA) performance.
\end{itemize}

\section{Related Works}

In processing documents, traditional approaches for feature identification generate static embeddings that do not contain contextual information. Methods such as Term Frequency - Inverse Document Frequency (TF-IDF) \cite{Jurafsky21}, word2vec \cite{Mikolov2013EfficientEO}, and GloVe \cite{Pennington2014GloVeGV} belong in this category. They generate numerical vector representations that contain some semantic information, but strictly speaking, are not contextual embeddings. Recent approaches construct contextual embeddings that represent a word in view of its context. LMs, such as BERT \cite{BERT} and GPT \cite{openai2024gpt4} belong in this category. These approaches can learn different representations for a word according to its surrounding context. The challenge with LMs, however, for using them on long financial documents, such as corporate earnings call transcripts, is the difficulty to extract document-level features, since the maximum number of word tokens these transformer LMs can handle is limited, and even if they can handle longer context windows, getting effective document-level representations still poses a big challenge.

On the semantic graph side, to our knowledge, AMR-based approaches have not been applied for long financial document classification tasks, such as earnings calls that can exceed 7,000 words in length. However, there have been several methods in different domains that utilize AMR for textual analysis. For example, researchers have used it for text classification \cite{shou-etal-2022-amr}, event detection \cite{li2015improving}, profanity and toxic content detection \cite{elbasani_kim_2022}, paraphrasability prediction and paraphrase generation \cite{lee-etal-2022-unsupervised}, and machine translation \cite{li-flanigan-2022-improving}. All of these utilize only sentence-level AMR, which is unsuitable for our purpose.

Methods for AMR parsing, which is the process of transforming text into AMR graphs, are well-studied. Transition-based parsers, such as \cite{zhou-etal-2021-structure} and \cite{lee-etal-2022-maximum}, provide SOTA sentence-level results, and AMR aligners, such as \cite{drozdov-etal-2022-inducing}, provide reliable AMR-to-text alignments that link each node entity to its corresponding word in the original sentence. There have also been recent works on parsing multi-sentence AMRs to preserve cross-sentence information. O'Gorman et al. \cite{ogorman-etal-2018-amr} provided a corpus of annotated multi-sentence AMRs, which was used by Naseem et al. \cite{naseem-etal-2022-docamr} to implement a new approach to constructing multi-sentence or document-level AMR representations. Since their approach is still limited to short documents (e.g., averaging about 429 words per document), it is unsuitable for our purpose. Instead, we transform the AMR graphs into a document-level graph specially designed for long documents.

As for utilizing document graphs and GNNs to perform graph learning for text classification in the finance domain, Medya et al. \cite{medya-earnings} designed and implemented StockGNN, which constructs document graphs based on the contextual window of each unique word in the document, applies a Gated GNN \cite{GGNN} on the document graphs, and concatenates the final embeddings of the document graphs with their respective doc2vec \cite{doc2vec} embeddings to generate the final document representation for text classification. They collected a corpus of general domain earnings calls from several sectors of the economy and predicted the financial impact of earnings calls on stock price using StockGNN. We use StockGNN as a baseline in our experiments. Importantly, StockGNN graphs are localized context graphs, and do not model semantics as we do via AMR.

\section{The FLAG Approach}
\begin{figure}[!ht]
\centering
  \includegraphics[scale=0.44]{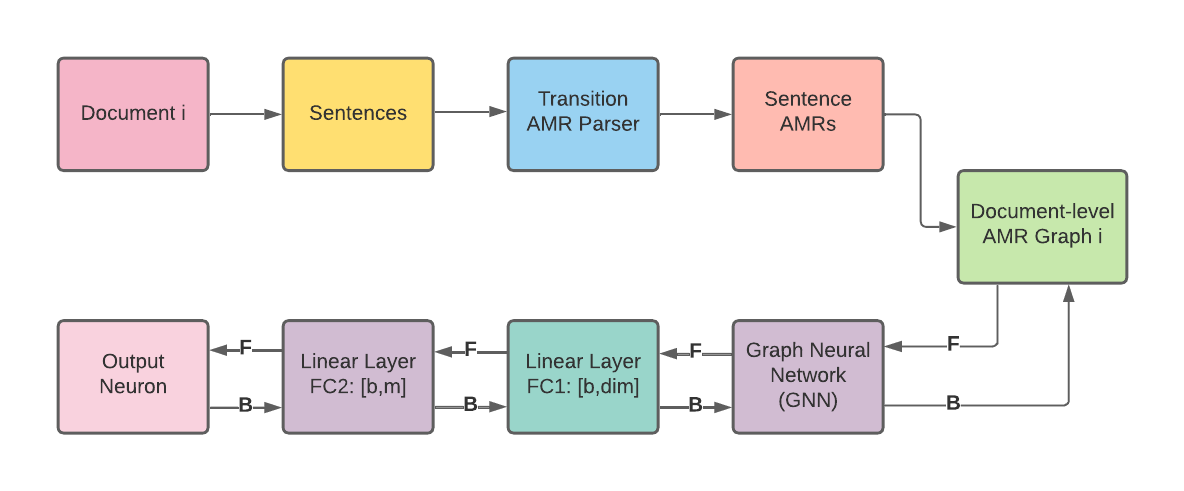}
  \caption{FLAG Architecture: Each document is parsed into sentences, which are converted into sentence AMR graphs. Using our hierarchical approach, we combine them into the document-level graph, which is endowed with the word embeddings and we then apply the GNN model to generate the final document virtual node embedding. This embedding is then passed through two fully connected linear layers to predict the target output. F: forward propagation; B: backward propagation.}
  \label{fig:pipeline}
\end{figure}

FLAG is an AMR-based GNN graph learning framework based on LM embeddings for long financial document classification. For each document, we parse all its sentences into AMR graphs, and construct a document-level AMR graph hierarchically with a sentence virtual node for every sentence and a document-level node to represent the document. We initialize each node with the word embedding of its corresponding word, generated from contextual LMs. Next, we apply a GNN model to generate the final document representations by taking the document virtual node embeddings. As such, our approach can be split into three phases: (1) Sentence AMR Parsing, (2) Document-level Graph Construction, and (3) GNN Model Training and Fine-Tuning, with the architecture illustrated in Figure~\ref{fig:pipeline}. Each of the phases is discussed next. 

\subsection{Sentence AMR Parsing}
In a corpus consisting of $N$ documents, $L = \{d_1, d_2, \cdots, d_N\}$, let $d_i$ represent the $i$-th document of the corpus. For a document $d_i$, we first sentencize the document into sentences, using the pre-trained Punkt tokenizer for English in the NLTK package \cite{bird-loper-2004-nltk}. This provides a sequence of sentences $S = \{s_1, s_2, \cdots, s_{m_i}\}$, where $m_i$ denotes the total number of sentences in $d_i$. Each sentence in $S$ is parsed into a sentence-level AMR graph using the transition AMR parser \cite{naseem-etal-2019-rewarding}, thereby generating a sequence of sentence graphs, $SG_i = \{sg_1, sg_2, \cdots, sg_{m_i}\}$.

\begin{figure}[!ht]
\centering
\includegraphics[scale=0.18]{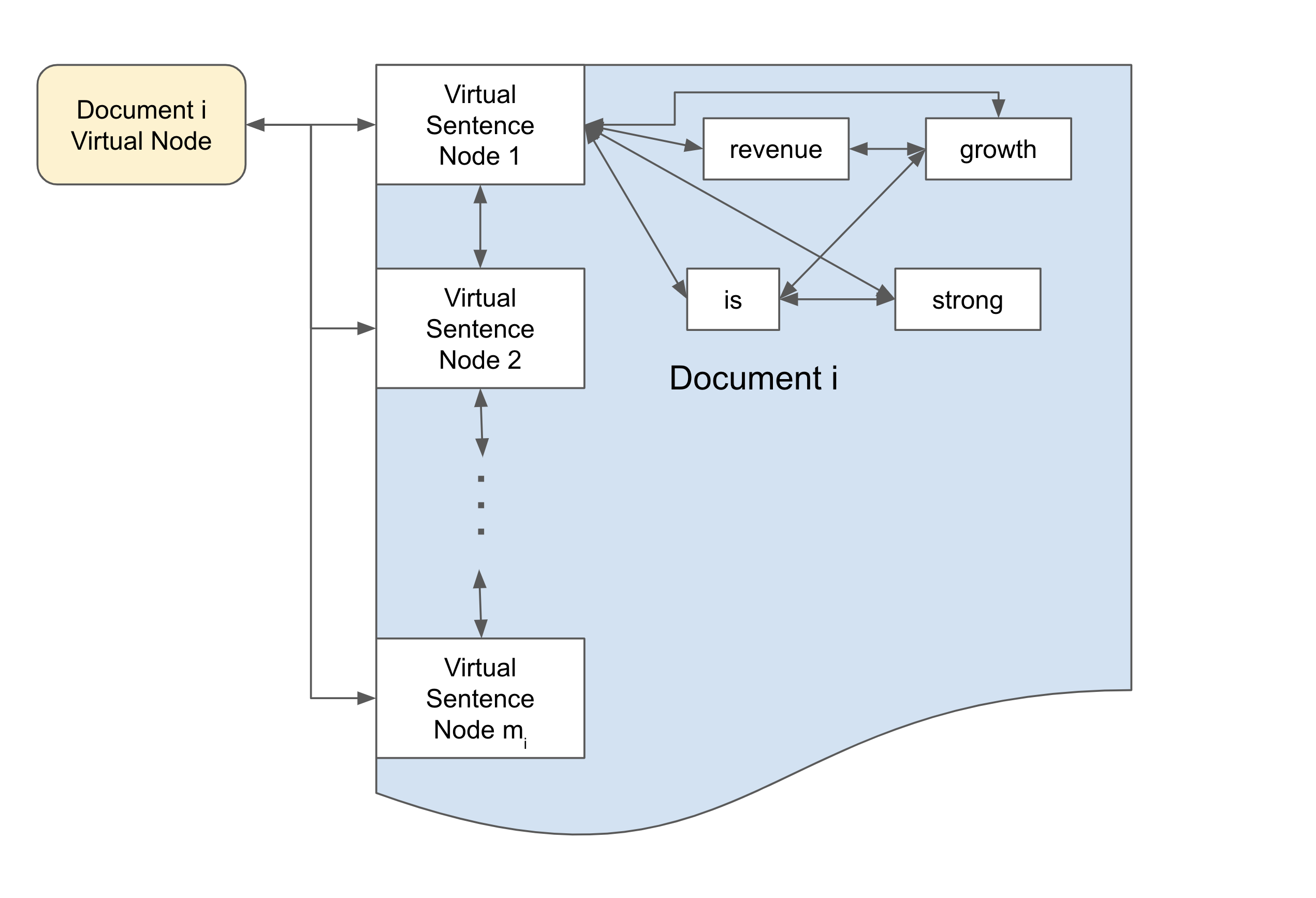}
  \caption{Document-level graph construction.}
  \label{fig:plan-C}
\end{figure}

\subsection{Document-level Graph Construction}\label{subsec:graph-construction}
Figure \ref{fig:plan-C} shows our document-level graph construction. For $d_i$, we have a sequence of sentence AMR graphs $SG_i = \{sg_1, sg_2, \cdots, sg_{m_i}\}$. For each sentence graph in $SG_i$, we make a virtual sentence node that connects to all the nodes in the sentence graph, culminating in a consecutive sequence of virtual sentence nodes $SN_i = \{sn_1, sn_2, \cdots, sn_{m_i}\}$, and then we connect these sentence virtual nodes in a consecutive manner, such that $sn_1$ is connected to $sn_2$, $sn_2$ is connected to $sn_3$, and so on, both forming a hierarchical representation of each sentence and preserving the order of sentences in the document. In turn, each sentence node in $SN_i$ is connected to a virtual document node $dn_i$, representing $d_i$. Overall, all the original nodes and edges in $SG_i$, all the virtual sentence nodes and their edges in $SN_i$, and the virtual document node $dn_i$ and its edges, form the graph structure $g_i$ for document $d_i$. We make every edge of the graph bidirectional, so that information can flow both ways during model training.

After constructing the graphs, we initialize the nodes with embeddings of their corresponding words in the original text, generated from the base LM method. We are able to do so, because the transition AMR parser \cite{drozdov-etal-2022-inducing} provides us with alignment between node entities in the sentence AMR graphs in $SG_i$ and their corresponding words in the original text of the sentences. 


The LM method that we chose is FinBERT \cite{yang2020finbert}, a specialized LM in the finance domain which generates token embeddings of size 768. We initialize the non-virtual nodes in our document-level graphs with the word embeddings of the original word in the sentence that they are aligned with. We pass each sentence of the document through the LM. A word's embedding is the average of the sum of the last 4 hidden state embeddings for each of the sub-tokens. Through this process, each non-virtual node in $g_i$ is initialized with word embeddings of size 768. All virtual nodes, $dn_i$ and all the nodes in $SN_i$, are initialized as zero vectors. 

\subsection{GNN Model Training and Fine-Tuning}
\label{subsec:gnn-train}
Given document-level graphs constructed for every document in the corpus $L$, forming a collection of graphs, $G_L = \{g_1, g_2, \cdots, g_N\}$, we apply an initial MLP layer on the graph features to transform the original node embedding dimension of 768 into another hidden dimension. We then train a GNN model on the transformed graphs, specifically, GATv2 \cite{GATv2}, an attention-based GNN model which has been theoretically proven to achieve dynamic attention. Afterwards, we take the embedding vector of the virtual document node $dn_i$ as the document representation for the graph $g_i$, denoted $\mathbf{h_{i}}$. We then use a linear layer:
\begin{equation}
	\mathbf{x_i} = W_{1}\mathbf{h_i} + \mathbf{b_{1}},
\end{equation}
followed by another linear layer that outputs the final predicted one-hot label, $\mathbf{\hat{y_{i}}}$, after a softmax.
\begin{equation}
	\mathbf{\hat{y_{i}}} = softmax(W_{2}\mathbf{x_i} + \mathbf{b_{2}})
\end{equation}
Finally, we use the cross entropy loss, $\mathcal{L} = -\sum_{k}y_{k}\log(\hat{y_{k}})$, where $\hat{y_{k}}$ is the $k$-th element of the predicted label, and $y_{k}$ is the $k$-th element of the target one-hot label.

\section{Empirical Evaluation}

We now present experimental results to evaluate the efficacy of FLAG. We utilized the transition AMR parser \cite{naseem-etal-2019-rewarding} with the AMR 2.0 structured BART large model pre-trained checkpoint \cite{drozdov-etal-2022-inducing} to perform AMR parsing and alignment, and we used the Deep Graph Library (DGL) \cite{wang2019dgl} for graph construction and initialization, as well as for GNN model training. 

\begin{table}[!ht]
 \centering
 \footnotesize
 \caption{Best hyperparameters for all methods: (e) Number of epochs, and (lr) learning rate.} 
 \label{tab:hyperparameters}
\begin{tabular}{|l|l|l|l|l|l|}
\hline
                         & Finance      & Health   & Materials      & Service  & Tech     \\\hline
FinBERT (E)          & 7        & 11       & 7        & 9        & 3        \\
FinBERT (LR)  & $10^{-5}$ & $10^{-6}$ & $10^{-5}$ & $10^{-6}$ & $10^{-5}$ \\\hline
StockGNN (E)         & 717      & 725      & 316      & 178      & 187      \\
StockGNN (LR)  & $10^{-5}$ & $10^{-5}$ & $10^{-5}$ & $10^{-5}$ & $10^{-5}$ \\\hline
FLAG (E)             & 14       & 9       & 17       & 10       & 15       \\
FLAG (LR)  & $10^{-5}$ & $10^{-5}$ & $10^{-5}$ & $10^{-5}$ & $10^{-5}$ \\\hline
\end{tabular}

\end{table}

For the hyperparameter tuning for FLAG, we train the framework for up to 20 epochs; for LM baselines, we train up to 30 epochs; and for StockGNN and related methods, we train up to 1000 epochs. We vary the learning rate from $10^{-2}$ to $10^{-6}$, and select the model with the lowest validation error as the best model. The best hyperparameters for each method are reported in Table \ref{tab:hyperparameters}. Our code and datasets are publicly available on github via
\url{https://github.com/Namir0806/FLAG}.

\subsection{Dataset and Metrics}
\label{subsec:dataset-and-metrics}

We take corporate earnings call transcript data as the subject of our analysis. An earnings call is a conference call between company executives and the financial community. It is usually held on a quarterly basis following the release of a company's earnings report. On this call, the management reviews the company's performance for a specific period, as well as potential risks and future plans, which can cause subsequent stock prices to shift dynamically~\cite{chen_2021}.

To evaluate FLAG, we compare its performance with baseline methods on both the Medya et al. earnings call dataset \cite{medya-earnings}, which contains earnings calls from 5 sectors of the economy during period from 2010 to 2019, and a new dataset, the S\&P 1500 earnings call corpus, which contains earnings calls of companies from the \href{https://www.spglobal.com/spdji/en/indices/equity/sp-composite-1500/}{S\&P 1500 Composite Index} during the period from 2010 to 2023, that we collected.

\begin{table}[!ht]
 \centering
 \caption{Medya Earnings Calls Dataset: $N$ is number of documents and $L$ is average document length.}
 \label{tab:medya-call-stats}
\begin{tabular}{|l|r|r|r|r|}
\hline
                & Train/Val $N$           & Test $N$            & Train/Val $L$                          & Test $L$                          \\\hline
Finance         & 14930           & 2036            & 7351.20                        & 6668.94                        \\\hline
Health          & 9869            & 1646            & 7083.55                        & 6591.70                        \\\hline
Materials & 9240            & 1256            & 7165.85                        & 6452.11                        \\\hline
Service         & 14890           & 1983            & 7498.45                        & 6928.90                        \\\hline
Technology      & 14319           & 2069            & 7112.78                        & 6785.09             \\\hline
\end{tabular}
\end{table}
\begin{table}[!ht]
\centering
\caption{FLAG graph statistics for the Medya earnings call dataset: N is number of nodes, E is number of edges, and D is degree.}
\begin{tabular}{|l|l|l|l|}
\hline\
        & Avg. N & Avg. E & Avg. D\\\hline
{\bf Finance} (Train/Val)     & 6307.34          & 19559.83         & 3.10\\
(Test) & 5784.46          & 17921.15         & 3.10               \\\hline
{\bf Health} (Train/Val)  & 6161.94          & 19115.60         & 3.10\\
(Test) & 5775.57          & 17908.75         & 3.10               \\\hline
{\bf Materials} (Train/Val)    & 6261.34          & 19400.57         & 3.10\\
(Test) & 5640.78          & 17464.83         & 3.10               \\\hline
{\bf Service} (Train/Val) & 6464.74          & 20066.69         & 3.10\\
(Test) & 5993.24          & 18598.27         & 3.10               \\\hline
{\bf Tech} (Train/Val)   & 6192.78          & 19215.09         & 3.10\\
(Test) & 5934.40          & 18407.39         & 3.10              \\\hline
\end{tabular}
\label{tab:medya-flag-graph-stats}
\end{table}

\medskip
\noindent
\textbf{Medya Earnings Call Dataset}
The Medya et al. \cite{medya-earnings} earnings call dataset consists of earnings calls during the 2010 to 2019 period. It is split into five sectors: finance, health, basic materials, service, and technology. They used this dataset to predict a binary trend label, which they call value-based labels. They define the label function $y_{v}(T_{d}^{c})$ for an earnings call transcript $T_{d}^{c}$ of a company $c$ on the day $d$ as follows:
\begin{equation}
y_{v}(T_{d}^{c}) = \begin{cases} 1, & if S_{d+1}^{c} > S_{d-1}^{c} \\
0, & otherwise,
\end{cases}
\end{equation}
where $S_{d+1}^{c}$ and $S_{d-1}^{c}$ denote the closing stock prices of company $c$ on the following and previous business day respective to day $d$.

Since this task is to analyze the trend, it aims to capture the immediate financial impact of an earnings call, so the target is daily and more granular than most financial impact analyses in industry. Moreover, in the real world, financial impact analyses in the context of portfolio management or asset pricing involve multiple complex factors, both qualitative and quantitative, and the textual signals from earnings calls only form a fraction of all the factors that can affect the stock price trends. Therefore, the purpose of experimenting on this dataset in predicting the daily value-based label is to evaluate the effectiveness of document representations produced by different methods 
for predicting immediate price movements only from textual data.

All the earnings calls during the period of 2010 to 2018 are used as the training/validation set (with a 80:20 split), and all the earnings calls during 2019 are used as the test set, as is done by Medya et al. \cite{medya-earnings}. The detailed data statistics are listed in Table \ref{tab:medya-call-stats}. As for the document-level graphs constructed using the FLAG approach from earnings calls for the Medya dataset, the detailed statistics are shown in Table \ref{tab:medya-flag-graph-stats}.

\begin{table}[!ht]
 \centering
 \caption{S\&P 1500 Earnings Calls Dataset statistics, with the total number of documents ($N$) and average document length ($L$).}
 \label{tab:sp1500-call-stats}
\begin{tabular}{|l|r|r|}
\hline
                    & $N$ & $L$ \\\hline
Train/Val & 55696               & 8138.38             \\\hline
Test             & 6049                & 7912.46             \\\hline
\end{tabular}
\end{table}

\begin{table}[!ht]
\centering
\caption{FLAG graph statistics for the S\&P 1500 earnings call dataset.}
\label{tab:sp1500-flag-graph-stats}
\begin{tabular}{|l|c|c|c|}
\hline
                 & Avg. \# of nodes & Avg. \# of edges                        & Avg. degree\\\hline
Train/Val & 7093.21          & 28883.19                                & 4.07\\
Test & 6512.83          & 26353.94                    & 4.05                \\\hline
\end{tabular}
\end{table}

\medskip\noindent
\textbf{S\&P 1500 Earnings Call Corpus}
The S\&P 1500 earnings call corpus dataset we collected contains more recent data, and consists of earnings call transcripts from the \href{https://www.spglobal.com/spdji/en/indices/equity/sp-composite-1500/}{S\&P 1500 Composite Index} for the period 2010 to 2023. It includes companies from all sectors and represents the overall U.S. equity market. Unlike for the Medya dataset, we use this dataset to predict weekly value-based labels. For this dataset, we define the label function $y_{v}(T_{d}^{c})$ for an earnings call transcript $T_{d}^{c}$ of a company $c$ on day $d$ as follows:
\begin{equation}
y_{v}(T_{d}^{c}) = \begin{cases} 1, & if \frac{1}{5}\sum_{i=1}^{5}S_{d+i}^{c} > \frac{1}{5}\sum_{i=1}^{5}S_{d-i}^{c} \\
0, & otherwise,
\end{cases}
\end{equation}
where $\frac{1}{5}\sum_{i=1}^{5}S_{d+i}^{c}$ and $\frac{1}{5}\sum_{i=1}^{5}S_{d-i}^{c}$ denote the average value of the closing stock prices of company $c$ from the week preceding and after respectively of day $d$ (a business week is defined as 5 working days).

The target labels in this case are on a weekly basis, so it is closer to what an analyst would infer from a company's earnings calls about the future direction of the company in the following week. The purpose is to investigate the effectiveness of documents representations produced by different methods, but in the context of predicting a longer-term (weekly) target label.

All the earnings calls during the period of 2010 to 2021 are used as the training/validation set (with a 90:10 split), and all the earnings calls during 2022-2023 are used as the test set. The detailed statistics for the corpus are shown in Table \ref{tab:sp1500-call-stats}. The document-level graphs are constructed using the FLAG approach from earnings calls in the S\&P 1500 corpus, with the detailed statistics shown in Table \ref{tab:sp1500-flag-graph-stats}.

\subsection{Methods}
\label{subsec:methods}
We describe the baselines and FLAG used in our experiments.
\begin{itemize}
\item {\bf FinBERT \cite{yang2020finbert}}:
We compare with a baseline LM model, using a domain-specific model pre-trained on financial corpora, FinBERT \cite{yang2020finbert}. This baseline method truncates the long document to the maximum length that the LM can take, and applies the LM to generate document embeddings, which is then passed through two linear layers to predict the target. Overall, this serves as a representative transformer-based LM baseline.

\item {\bf StockGNN \cite{medya-earnings}}:
StockGNN \cite{medya-earnings} also utilizes contextual graphs and the Gated GNN (GGNN) method \cite{GGNN} to learn from long financial documents. However, it uses only a local contextual graph and does not leverage deeper semantics. Overall, this serves as a representative previous SOTA baseline method.

\item {\bf FLAG}:
For the GNN model in FLAG, we configure GATv2 with 4 layers, 8 attention heads per layer, and 512 as the hidden dimension that node embeddings of the original graphs are transformed to before applying the GNN model. We found this configuration to perform best empirically, which we discuss more in the ablation studies below. Each GATv2 layer reaches a node's immediate neighbours. Therefore, 4 layers are sufficient to reach from one node to all other nodes, as the diameter of the document graphs is close to 4.

\item {\bf GGNN on FLAG Graphs}: Since StockGNN uses only contextual graphs, we examine the effectiveness of GGNN~\cite{GGNN} on the AMR graphs constructed in FLAG. This baseline serves to showcase the added benefits from a more semantics-based AMR graph with the same GNN as used in StockGNN.
\end{itemize}

\subsection{Comparative Results}
For experiments, we train the models to minimize the training loss, and because the target value is a trend metric, we evaluate the quality of the binary trend label predictions by evaluating the accuracy, precision (macro average), and recall (macro average) of the predictions, as is done in Medya et al. \cite{medya-earnings}, as well as the F1 score.

\begin{table}[!ht]
\centering
\caption{Experiment results on the Medya earnings call dataset. Best results are marked in bold.}
\small
\tabcolsep=0.11cm
\begin{tabular}{|l|ccccc|}
\hline
         & &&{\bf Accuracy}&&       \\\hline  
         & Finance            & Health         & Materials            & Service        & Tech          \\\hline 
FinBERT  & 0.574          & 0.547          & 0.516          & 0.496          & 0.541          \\
StockGNN & 0.559          & 0.584          & 0.561          & 0.554          & 0.551\\
FLAG     & \textbf{0.619} & \textbf{0.614} & \textbf{0.597} & \textbf{0.626} & \textbf{0.637}\\\hline
         & &&{\bf Precision}  &&  \\\hline 
         & Finance            & Health         & Materials            & Service        & Tech          \\\hline 
FinBERT & 0.581          & 0.544          & 0.524          & 0.503          & 0.545\\
StockGNN & 0.563          & 0.583          & 0.562          & 0.554          & 0.557\\
FLAG  & \textbf{0.615} & \textbf{0.616} & \textbf{0.595} & \textbf{0.629} & \textbf{0.641}\\\hline 
         & &&{\bf Recall}&&   \\\hline
& Finance            & Health         & Materials            & Service        & Tech           \\\hline
FinBERT & 0.521          & 0.543          & 0.523          & 0.501          & 0.543          \\
StockGNN & 0.564          & 0.583          & 0.562          & 0.553          & 0.553          \\
FLAG  & \textbf{0.616} & \textbf{0.604} & \textbf{0.593} & \textbf{0.627} & \textbf{0.638} \\\hline
         &                &                & F1 Score       &                &                \\\hline
         & Finance        & Health         & Materials      & Service        & Tech           \\\hline
FinBERT  & 0.437          & 0.542          & 0.513          & 0.423          & 0.536          \\
StockGNN & 0.559          & 0.583          & 0.561          & 0.552          & 0.543          \\
FLAG     & \textbf{0.615} & \textbf{0.598} & \textbf{0.592} & \textbf{0.624} & \textbf{0.635} \\\hline
\end{tabular}

\label{tab:medya-dataset-experiment-results}
\end{table}



\medskip \noindent
\textbf{Results on the Medya Dataset}:\label{subsubsec:earnings-experiments}
Table \ref{tab:medya-dataset-experiment-results} shows the detailed results of our comprehensive experiment across all 5 sectors in the Medya et al. earnings calls dataset. As we can see, FLAG outperforms both FinBERT and StockGNN across all sectors of the economy. However, there are different degrees to how well FLAG performs in different sectors. Especially compared to the FinBERT baseline, we see the smallest improvement in performance in the financial sector. Overall, the highest absolute performance is in the technology sector. Therefore, in the spirit of fintech, we take the financial sector and the technnology sector as subjects of analyses in both the ablation studies and the case study below.

With regards to FinBERT, FLAG is able to achieve significant performance gains over directly applying the domain-specific LM. Especially in the service and technology sectors, we see very high performance gains with FLAG (e.g., 26.2\% gain for service), and in the financial sector, it outperforms FinBERT, however by a lesser margin (7.8\% gain). We find the same trend for performance using StockGNN. The improvement in the service and technology sector is generally higher than in the other sectors. We posit that this trend is due to the differences in the nature of earnings calls in different sectors of the economy and how the stock market reacts to an earnings call event in each particular sector, which we analyze in the case study below.

Earnings calls have an immediate financial impact on stock prices, and through this set of experiments, we show that FLAG is able to capture that better. In other words, as a more semantically meaningful approach, it is able to achieve better performance in predicting stock price movement trends for which the input document is a main contributing factor. This indicates that choosing the sparse connections in a semantically meaningful way, as done via the AMR graphs, helps the model achieve better performance in trend analysis. We validate this conclusion through an ablation study of applying Gated GNN, the GNN used in StockGNN, on the AMR document-level graphs (discussed below). This finding is important for building a textual element within asset pricing model in the real world, where the textual signals implications for stock price trends must be understood.

\begin{table}[!ht]
\centering
\small
\caption{Experiment results on the S\&P 1500 earnings call dataset. Best results are marked in bold.}
\begin{tabular}{|l|c|c|c|c|}
\hline
         & Accuracy       & Precision      & Recall         & F1 Score       \\\hline
FinBERT  & 0.549          & 0.543          & 0.530          & 0.501          \\
StockGNN & 0.556          & 0.551          & 0.543          & 0.533          \\
FLAG     & \textbf{0.598} & \textbf{0.598} & \textbf{0.586} & \textbf{0.578}\\\hline
\end{tabular}
\label{tab:sp1500-dataset-experiment-results}
\end{table}

\medskip\noindent
\textbf{Results on S\&P 1500 Dataset}:
Table \ref{tab:sp1500-dataset-experiment-results} shows the experiment results on the S\&P 1500 dataset, in the context of predicting weekly average price trends. Once again, FLAG performs better in predicting longer-term price trend metrics at the weekly granularity when compared with FinBERT and StockGNN, with the graph-based approach of StockGNN performing better than FinBERT. This corpus consists of earnings calls from all sectors in the U.S. equity market, as represented in the index. As such, the absolute performance gains of FLAG on this corpus are lower compared to the sector-specific Medya earnings call dataset, albeit the target metric scope is different: the latter is on a daily granularity. Overall, the results show that not only are there indications for weekly stock price trends contained in the soft information of the earnings calls, but that FLAG is able to extract this better with its semantically meaningful graph representations of documents, even across various different sectors of the economy.

\begin{table}[!ht]
\centering
\caption{Performance of StockGNN vs. Gated GNN with FLAG.}
\label{tab:ggnn-on-flag-graph-ablation}
\begin{tabular}{|l|cccc|}
\hline
                    &                & \multicolumn{2}{c}{Finance}         &                \\\hline
                    & Accuracy       & Precision      & Recall         & F1 Score       \\\hline
StockGNN            & 0.541          & 0.513          & 0.509          & 0.489          \\
GGNN on FLAG graphs & \textbf{0.547} & \textbf{0.540} & \textbf{0.540} & \textbf{0.540} \\\hline
                    &                & \multicolumn{2}{c}{Tech}        &                \\\hline
                    & Accuracy       & Precision      & Recall         & F1 Score       \\\hline
StockGNN            & 0.560          & 0.560          & 0.560          & 0.560          \\
GGNN on FLAG graphs & \textbf{0.573} & \textbf{0.574} & \textbf{0.574} & \textbf{0.573}\\\hline
\end{tabular}
\end{table}

\subsection{Ablation Studies}\label{subsubsec:ablation}

\paragraph{Efficacy of AMR document-level graphs: a comparison between StockGNN and Gated GNN on FLAG graphs}

As FLAG and StockGNN utilize different GNNs, with FLAG using GATv2 and StockGNN using Gated GNN, it is of interest to evaluate the performance of document graphs constructed with FLAG versus document graphs constructed with StockGNN. Therefore, we apply the same configurations of GGNN that StockGNN uses, but on the documents graphs constructed with FLAG, and compare its performance against StockGNN generated graphs, on the financial and technology sectors of the Medya dataset. We ran StockGNN for 100 epochs and GGNN on FLAG graphs for 50 epochs. Table \ref{tab:ggnn-on-flag-graph-ablation} shows the results.

For both the sectors, there is value added in using FLAG-based document graphs, even without using an attention-based GNN, such as GATv2. We see the same performance trends as the main experiments conducted on the Medya dataset, where the performance improvement is less for the financial sector, compared to the technology sector. We delve a bit into why this may be the case in the case study below.

\begin{table}[!ht]
\centering
\caption{Ablation of various different GNNs applied on FLAG graphs.}
\label{tab:gnn-ablations}
\begin{tabular}{|l|cccc|}
\hline
     &                & \multicolumn{2}{c}{Finance}         &                \\\hline
     & Accuracy       & Precision      & Recall         & F1 Score       \\\hline
FLAG & \textbf{0.619} & \textbf{0.615} & \textbf{0.616} & \textbf{0.615} \\
GAT  & 0.603          & 0.598          & 0.599          & 0.598          \\
GCN  & 0.565          & 0.553          & 0.551          & 0.550          \\
GGNN & 0.547          & 0.540          & 0.540          & 0.540          \\
PNA  & 0.561          & 0.281          & 0.500          & 0.360          \\\hline
     &                & \multicolumn{2}{c}{Tech}        &                \\\hline
     & Accuracy       & Precision      & Recall         & F1 Score       \\\hline
FLAG & \textbf{0.637} & \textbf{0.641} & \textbf{0.638} & \textbf{0.635} \\
GAT  & 0.619          & 0.627          & 0.621          & 0.614          \\
GCN  & 0.594          & 0.595          & 0.595          & 0.594          \\
GGNN & 0.573          & 0.574          & 0.574          & 0.573          \\
PNA  & 0.511          & 0.525          & 0.516          & 0.462  \\\hline
\end{tabular}
\end{table}

\paragraph{Efficacy of attention-based GATv2: a comparison of different GNNs applied on FLAG graphs}

Coupling dynamic attention offered by GATv2 with the structure of FLAG-based document graphs is an important aspect of the FLAG methodology and has been shown to achieve superior performance. On account of the large size of the document graphs produced by FLAG, as shown in the dataset metrics in Section \ref{subsec:dataset-and-metrics}, graph transformers cannot be applied on FLAG graphs. However, there are other graph convolution networks that can be applied to FLAG graphs, including conventional GCN, PNA \cite{corso2020pna}, GAT \cite{velickovic2018graphGAT}, which does not have dynamic attention, and Gated GNN \cite{GGNN} (also compared above). We experiment with these GNNs and present the results in Table \ref{tab:gnn-ablations}. We find that GATv2 coupled with FLAG graphs offers an edge in performance that other non-attention-based GNNs or GNNs without dynamic attention cannot match. Also interesting is that the GCN model also outperforms GGNN and PNA on the earnings call graphs.

\begin{table}[!ht]
\caption{Ablation of FLAG on the Financial Sector.}
\label{tab:flag-on-medya-fin-ablations}
\centering
\begin{tabular}{|l|llll|}
\hline
\# Layers & Accuracy & Precision & Recall & F1 Score\\
\hline
& \multicolumn{4}{c|}{\bf 4 attention heads, 256dim}\\
\hline
1 & 0.561          & 0.281          & 0.500          & 0.360          \\
2 & 0.604          & 0.600          & 0.601          & 0.600          \\
3 & 0.569          & 0.588          & 0.584          & 0.568          \\
4 & 0.596          & 0.598          & 0.599          & 0.595          \\
5 & 0.603          & 0.598          & 0.599          & 0.599          \\
6 & 0.605          & 0.596          & 0.591          & 0.590          \\
\hline
& \multicolumn{4}{c|}{\bf 8 attention heads, 512dim}\\
\hline
1 & 0.561          & 0.281          & 0.500          & 0.360          \\
2 & 0.602          & 0.598          & 0.600          & 0.598          \\
3 & 0.611          & 0.607          & 0.608          & 0.607          \\
4 & \textbf{0.619} & \textbf{0.615} & \textbf{0.616} & \textbf{0.615} \\
5 & 0.604          & 0.598          & 0.598          & 0.598          \\
6 & 0.598          & 0.597          & 0.598          & 0.596          \\
\hline
& \multicolumn{4}{c|}{\bf 12 attention heads, 768dim}\\
\hline
1 & 0.561          & 0.281          & 0.500          & 0.360          \\
2 & 0.595          & 0.597          & 0.598          & 0.594          \\
3 & 0.585          & 0.590          & 0.591          & 0.585          \\
4 & 0.616          & 0.608          & 0.606          & 0.607          \\
5 & 0.571          & 0.555          & 0.546          & 0.537          \\
6 & 0.565          & 0.550          & 0.510          & 0.412         \\
\hline
\end{tabular}
\end{table}

\begin{table*}[!ht]
\centering
\caption{A juxtaposition of sentences that FLAG deemed to be important from the worst performing document in the financial sector (Cushman \& Wakefield, Q2, 2019) and the best performing document in the technology sector (Paycom, Q1, 2019).}
\label{tab:case-study}
\begin{tabular}{|p{7cm}|p{7cm}|}
\hline
Cushman \& Wakefield, Q2, 2019                                                                                                                                       & Paycom, Q1, 2019                                                                                                                                                                         \\\hline
This year marks 20 years of H R Block being a Cushman Wakefield client.                                                                                      & Our revenue growth continues to be primarily driven by new business wins.                                                                                                     \\\hline
These are among the primary metrics we monitor to assess commercial real estate supply and demand and to provide a foundation for our business forecast. & For fiscal 2019, we are  increasing our revenue guidance to a range of 718 million to 720 million or   approximately 27 year over year growth at the midpoint of the range  \\\hline
Growth continues to be strong.                                                                                                                               & Well, first on the percentage   of those clients that have committed to a 400 employee usage strategy.                                                                       \\\hline
Obviously, last year we had very strong growth in all the courses in   leasing.                                                                             & We are seeing greater usage prior to implementation as we continue to   drive usage even prior to the full deployment go live of the system.                                  \\\hline
We expect to see some of those trends continue in the second half.                                                                                           & I would say the mix has been consistent as it's been in the past.                               \\\hline                                                                            
\end{tabular}
\end{table*}

\paragraph{Various configurations of GATv2: number of layers, attention heads, and different hidden dimensions}
GATv2 is a GNN that has many possible configurations, including the number of layers and attention heads, and different hidden dimensions of transformed input graphs (in FLAG). In our ablations, we experimented with the number of layers ranging from 1 to 6, the number of attention heads varying from 4, 8, and 12, and hidden dimensions of 256, 512, and 768. Table~\ref{tab:flag-on-medya-fin-ablations} shows the effect on the performance of frameworks with different numbers of GATv2 layers, as we vary the attention heads and the hidden dimensions. Results are shown for the case when the projected dimensionality for the attention is fixed as 64, be it $64=256/4$, or $64=512/8$, or $64=768/12$.  We find that the configuration with 4 layers, with 8 attention heads each, and a hidden dimension of 512 has consistently better performance across sectors, and that is chosen as the default configuration for FLAG.

\subsection{Case Study}\label{subsubsec:case-study}

As we have seen in the experiment results, FLAG achieves higher performance gains in sectors such as technology, but not as much in sectors like the financial sector. We posited that the stock market reacts differently to earnings calls in the financial sector versus the technology sector. To investigate this further, we applied the GNNexplainer \cite{GNNexplainer} method to the document on which FLAG performed the worst in the financial sector (for {\it Cushman \& Wakefield, Q2, 2019}) and on the document that FLAG performed the best in the technology sector (for {\it Paycom, Q1, 2019}), in terms of cross-entropy loss. We trained GNNexplainer with 3 hops for 1000 epochs, and generated the edge mask. This shows, in numerical terms, which edges in the FLAG graph are determined to be important in making the prediction. We then looked at all the edges going into the document virtual node from sentence virtual nodes, and examined the important edges to extract the top sentences that were considered to be more important for the final classification. Table \ref{tab:case-study} shows some example top sentences from the two documents.

In the case of Cushman \& Wakefield, the actual stock price went down, but FLAG predicted that it would go up. Looking at the top sentences, we can see why it made such a prediction, as the earnings call speaks about growth in the past and how they expect the growth to continue in the future. However, it would seem that the market did not believe Cushman \& Wakefield projections, and the stock price actually went down as a result. In the case of Paycom, the stock price went up, and FLAG predicted correctly that it would go up. From the top sentences FLAG identified, we see the company speaking optimistically about its growth and a new system that employees are using more and more, which the market reacted positively to, leading to an uptick in stock price.

Albeit only one representative case study, it suggests a potential trend where the market does not ``believe'' firms in the financial sector on the face value of statement made in earnings calls as much as they do firms in other sectors, such as the technology sector, where companies deal with more tangible assets. This warrants examining more on the broader characteristics of earnings calls in different sectors and how market reacts to them differently, and we plan to conduct this analysis in our future work.

\section{Conclusion and Future Works}
This work presents the first use of AMR-based graphs in a predictive framework for long document classification tasks, such as financial trend prediction. 
As demonstrated by the experiments, in predicting both immediate and longer-term price trends based on textual signals from earnings calls, FLAG with underlying FinBERT-generated embeddings is able to show SOTA performance against both LM and previous baselines. Therefore, our next step is to endow FLAG with embeddings extrapolated from more advanced LLMs such as FinGPT \cite{yang2023fingpt}. However, since unlike FinBERT, which is an encoder-decoder model, FinGPT is a decoder and causal model, it will require some modifications to generate contextually informed embeddings.

We have also seen the performance of FLAG varying from sector to sector, which warrants further analysis into the characteristics of earnings calls in different sectors and their interactions with the market. Specifically, we plan to identify parts of documents that the model considers as important to its prediction, leveraging the explainability offered by semantic document graphs and attention-based GNNs in the FLAG deep learning framework, in order to find particular patterns and scenarios in which correct or incorrect predictions are made, both in specific sectors and across all sectors.

Moreover, in the real-world use case scenario of price trend analysis, financial analysts generate numerical predictions from an aggregation of a variety of factors, and not only from textual signals. Therefore, we want to use FLAG to generate more complex qualitative insights from textual signals that analysts cannot get directly from the data at hand. To achieve this, we need to develop a retrieval methodology that works on document-level graphs instead of on blocks of text, which would enable generative models to leverage FLAG for better semantically informed insights. We also need further discussions and collaborations with stakeholders in the financial industry to identify which types of qualitative insights are of value.

\section*{Acknowledgement}
This work was supported by an industry funded award from the RPI-Stevens NSF IUCRC  Center for Research toward Advancing Financial Technologies (NSF Award \#: 2113850).

\bibliographystyle{IEEEtran}
\bibliography{main}

\end{document}